\documentclass[twocolumn,twoside,slac_two]{revtex4}
\usepackage{graphicx}
\usepackage{fancyhdr}
\begin{document}

\title{The XYZ's of $c\bar{c}$: Hints of Exotic New Mesons}

%

\author{Stephen Godfrey}
\affiliation{Ottawa Carleton Institute of Physics, 
Department of Physics, Carleton University, Ottawa K1S 5B6, Canada}

\begin{abstract}
I discuss the nature of the new charm and charmonium like states 
observed in the last few years and measurements that can test these 
assignments.  In particular it appears that the $X(3943)$ is the 
$\eta_c''(3^1S_0)$ which can be tested by looking for it in 
$\gamma\gamma \to D\bar{D}^*$, the $Y(3943)$ is the $\chi_{c1}'(2^3P_1)$ 
which can be tested by looking for it in $D\bar{D}$ and $D\bar{D}^*$, the 
$Z(3930)$ is the $\chi_{c2}'(2^3P_2)$ 
which can be confirmed by looking for it in $D\bar{D}^*$.  If the 
$X(3872)$ is confirmed to have $J^{PC}=1^{++}$ it is almost certainly 
a multiquark state while if its $J^{PC}$ is found to be $2^{-+}$ is is 
likely the $1^1D_2$ state.  The $Y(4260)$ appears to be an extra 
$1^{--}$ which is most easily explained as a charmonium hybrid.  This 
can be tested by looking the $DD_1$ final state.
\end{abstract}

\maketitle

\thispagestyle{fancy}


\section{Introduction}
The last few years have seen a phenomenal resurgence in charm 
and quarkonium spectroscopy. It began 
in July 2002 when CLEO presented evidence for 
a $D$-wave $b\bar{b}$ meson \cite{Bonvicini:2004yj}.  
This was the first new quarkonium state 
to be observed in almost twenty years.  Since then eight new 
charmonium like states have been observed 
\cite{Choi:2002na,Asner:2003wv,Rosner:2005ry,Abe:2005hd,Abe:2004zs,Uehara:2005qd,Aubert:2005rm,Aubert:2005zh,Coan:2006rv,Choi:2003ue,Acosta:2003zx,Abazov:2004kp,Aubert:2004ns}
plus the $B_c$ \cite{Abe:1998wi,Abe:1998fb,Acosta:2005us}, 
the puzzling 
$D_{sJ}^{(*)}$ states \cite{Aubert:2003fg,Besson:2003cp,Abe:2003jk,Evdokimov:2004iy}, 
and the broad $D$ $P$-wave states \cite{Abe:2003zm,Link:2003bd,Anderson:1999wn}.  
This 
collection of states have in some cases confirmed quark model and Lattice QCD 
calculations while in other cases challenged our understanding.  In 
other words it's an exciting time to be a 
spectroscopist!  In this mini-review I survey these new states, 
concentrating on conventional interpretations and suggesting 
non-quarkonium explanations when all else fails.  My talk is 
complemented by Voloshin's talk which concentrates on multiquark descriptions 
of some of these states \cite{Voloshin:2006wf}.

This mini-review starts with some brief remarks about conventional meson 
spectroscopy, radiative transitions, and strong decays along with 
comments about hybrid mesons -- states with an excited gluonic degree of 
freedom.  It is followed by a brief discussion of the charm-strange 
mesons and broad $P$-wave charm mesons primarily focusing on some 
recent experimental results and how we can test the identity of these 
states.  The bulk of this review concentrates on the various 
new charmonium like states; the $X$, $Y$, $Z$'s of the title. In the 
final section I 
summarize my conclusions about these states and suggest 
experimental tests of the interpretations.  Some of these topics have 
recently been reviewed by Swanson \cite{Swanson:2006st}.

\section{General Remarks on Spectroscopy}

This section is intended to be a reminder of the template we are using 
to test for ``exotic'' states.  We use the predictions of a 
specfic model \cite{Godfrey:1985xj}
as our template and briefly describe some details of 
this approach.  An extensive review of quarkonium physics, including 
other calculations and detailed references, 
is given in Ref.~\cite{Brambilla:2004wf}.

In quark potential models,
conventional meson quantum 
numbers are characterized by the $J^{PC}$ given by
$P=(-1)^{L+1}$ and $C=(-1)^{L+S}$ where $S$ is the total spin of the 
$q\bar{q}$ pair and the total angular momentum $J$ is found by adding $S$ 
to $L$, the orbital angular momentum of the quark antiquark pair.  To 
obtain the quarkonium spectrum one starts with a potential and solves 
the eigenvalue equation, Schrodinger equation or otherwise,
for orbital and radial excitations.  
The potential typically consists of a short distance 
Coulomb potential expected from one-gluon-exchange and a 
linear confining potential at large separation.  This phenomenological 
potential is in good agreement with the static quarkonium potential 
calculated using Lattice QCD.  

In addition to the spin-independent potential there are spin dependent 
interactions which are $(v/c)^2$ corrections.  They are found by 
assuming that the short distance one-gluon-exchange is 
a Lorentz vector interaction and the confinement piece is Lorentz 
scalar. This gives rise to multiplet splittings.  For example, the 
$J/\psi -\eta_c$ splitting is attributed to a short distance 
$\vec{S}_q\cdot\vec{S}_{\bar{q}}$ contact 
interaction arising from the one-gluon-exchange while the splitting of 
the $P$-wave $\chi_c$ states is due to spin 
orbit interactions arising from one-gluon-exchange and the relativistic 
Thomas precession piece in addition to 
a tensor spin-spin interaction. The recent measurement of the $h_c$ 
mass is an important validation of this picture.

The properties of meson states can be further tested by calculating 
electromagnetic and strong decays (and for that matter weak decays of 
stable states) and comparing them to experiment.  The calculation of 
radiative transitions are straightforward and are described in many 
places \cite{Kwo88}.  
For the strong decays we rely on the $^3P_0$ decay model 
which describes most strong decays reasonably well
\cite{Micu:1968mk,LeYaouanc:1972ae,Ackleh:1996yt,Barnes:1996ff,Barnes:2002mu,Blundell:1995ev}.  
Decays of 
charmonium states up to $\sim 4.6$~GeV were calculated by Barnes 
Godfrey and Swanson \cite{Barnes:2005pb} with similar results 
obtained by Eichten Lane and Quigg \cite{Eichten:2005ga}.  
These results can be used to test the properties of a newly 
discovered state to see if and where it fits into the expected 
charmonium spectroscopy.

In addition to the conventional quarkonium states other hadron states 
are expected.  Multiquark states have 
a larger quark content than the conventional 
$q\bar{q}$. The details are rather complicated and has spawned 
a subfield in the literature studying the different predictions for 
various possible configurations.  For example, in one extreme these 
multiquark states consist of tightly bound $q^2\bar{q}^2$ and are 
referred to as ``tetra-quarks'' while the 
other extreme consists of loosely bound mesons such as $D\bar{D}$ and 
are referred to as molecules.  I refer you to Voloshin's contribution
\cite{Voloshin:2006wf}.

The other type of exotic quarkonium state are the so-called hybrid 
mesons which have an excited gluonic degree of freedom. These are 
described by many different models and calculational schemes
\cite{Barnes:1995hc}.  The 
picture I prefer is analogous to molecular physics where the quarks 
move in adiabatic potentials arising from the gluons which can be 
compared to nuclei 
moving in the adiabatic potentials arising from the electrons in 
molecules.  The lowest adiabatic surface leads to the 
conventional quarkonium spectrum while the excited adiabatic surfaces 
are found by putting the quarks into more complicated colour 
configurations.  The adiabatic potentials have been calculated using 
Lattice QCD \cite{Morningstar:2000mf,Bali:2003jq}.  
In the flux tube model \cite{Isgur:1984bm} the lowest excited adiabatic 
surface corresponds to transverse excitations of the flux tube and 
leads to a doubly degenerate octet of the lowest mass hybrids with 
quantum numbers $J^{PC}=0^{+-}$, $0^{-+}$, $1^{+-}$, $1^{-+}$, 
$2^{+-}$, $2^{-+}$, $1^{++}$ and $1^{--}$. The $0^{+-}$, $1^{-+}$, $2^{+-}$
quantum numbers are not possible in the quark model and are referred 
to as exotic quantum numbers.  If observed, they would unambiguously 
signal the existence of unconventional states. 
Lattice QCD and most models predict the 
lowest charmonium hybrid state to be roughly 4.2~GeV in mass.

Charmonium hybrids can decay via electromagnetic transitions, hadronic 
transitions such as $\psi_g\to J/\psi +\pi \pi$, and to open charm 
final states 
like $\psi_g\to D^{(*,**)}\bar{D}^{(*,**)}$.  The partial widths have 
been calculated using many different models.  There are some
general properties that seem to be supported by most models and by recent 
lattice QCD calculations.  Nevertheless there are no experimental results 
against which to test these calculations so one should take these 
predictions with a grain of salt.  Two important decay modes are:
\begin{enumerate}
\item $\psi_g\to D^{(*,**)}\bar{D}^{(*,**)}$.  Most calculations 
predict that the $\psi_g$ should decay to a $P$-wave plus an $S$-wave 
meson.  In other words $D(L=0)+D^{**}(L=1)$ final states should 
dominate with vanishing partial widths for decays to $D\bar{D}$ and a small 
partial width to $D\bar{D}^*$.
\item $\psi_g \to (c\bar{c})(gg)\to (c\bar{c})+(\pi\pi, \; \eta 
\ldots)$ This mode offers the cleanest signature.  If the total width 
is small it could have a significant branching fraction.  A recent 
lattice QCD calculation finds that these decays are potentially quite 
large, ${\cal O}(10\hbox{ MeV})$ although it should be noted that the 
calculation was for $(b\bar{b})_g\to\chi_b S$ where $S$ is a light 
scalar meson\cite{McNeile:2002az}.
\end{enumerate}

\section{Some Other New States}
Before proceeding to the puzzles I was asked to review  I want to 
mention several other new states which have added to our understanding 
of meson spectroscopy.
\begin{description}
\item[$\Upsilon(1D)$] This state was first announced by CLEO at the 
2002 ICHEP conference  \cite{Bonvicini:2004yj}.  It's mass of $M(\Upsilon)=10161.1\pm 0.6 
(stat)\pm 1.6(syst)$~MeV
is in good agreement with potential models \cite{Godfrey:1985xj}
and lattice QCD calculations \cite{Gottlieb:2003bt}.
\item[$B_c$] While observed previously \cite{Abe:1998wi,Abe:1998fb}
the CDF collaboration recently 
presented a precise mass measurement which could confront theoretical 
predictions \cite{Acosta:2005us}.  The observed mass of 
$M(B_c)=6287.0\pm 4.8 (stat)\pm 1.1 (syst)$~MeV compares favourably to 
the lattice QCD result of $6304\pm 12$~MeV \cite{Allison:2004be}
and the quark potential model result of 6271~MeV 
\cite{Godfrey:1985xj,Godfrey:2004ya}. 
\item[$\eta_c'$] This state was recently observed by 
Belle \cite{Choi:2002na}
 and CLEO \cite{Asner:2003wv}.  The combined mass of 
$M(\eta_c')=3637.4\pm 4.4$~MeV is slightly higher than the quark model 
prediction of 3623~MeV \cite{Godfrey:1985xj,Barnes:2005pb}
so that the quark model slightly overestimates 
the $2^3S_1-2^1S_0$ splitting.  Eichten Lane and Quigg \cite{Eichten:2005ga}
studied the 
coupled channel contributions to $c\bar{c}$ states and found that this 
reduces the splitting, bringing it into better agreement with 
experiment.  
\item[$h_c$] This state was recently observed by the CLEO collaboration 
\cite{Rosner:2005ry}.  Its mass of $M(h_c)=3524.4\pm 0.6(stat)\pm 0.4 
(syst)$~MeV gives the $^3P_J-^1P_1$ splitting of 
$M(^3P_J)-M(^1P_1)=1.0\pm0.6(stat)\pm 0.4(syst)$~MeV which implies a 
very short range contact interaction supporting the Lorentz-vector 
1-gluon-exchange plus Lorentz-scalar linear confining potential.  The 
predictions for this splitting had a very large variation so this 
measurement is a useful constraint on models 
\cite{Godfrey:2005un,Godfrey:2002rp}.
\end{description}
Taken together these results provide an important test of quarkonium 
spectroscopy calculations and help calibrate the reliability of the
predictions.

\section{The $D_{sJ}(2317)$ and $D_{sJ}(2460)$}
The $D_{sJ}(2317)$ was first observed by Babar \cite{Aubert:2003fg} 
and the $D_{sJ}(2460)$ by CLEO \cite{Besson:2003cp}.  
Both were subsequently seen and studied by Belle \cite{Abe:2003jk}.  Their 
properties are consistent with $J^P=0^+$ and $1^+$ respectively.  Two 
broad $P$-wave charm-strange mesons were expected with the $J^P=0^+$
state decaying to $DK$ and the $1^+$ to $D^*K$ \cite{Godfrey:1986wj}.  
But both states are very narrow with the 
$D_{sJ}(2317)$ below the $DK$ threshold and the $D_{sJ}(2460)$ 
below the $D^*K$ threshold. This 
unexpected behavior created a major theory industry describing the 
$D_{sJ}$ states as multiquark states, molecular states, $D\pi$ atom, 
and as conventional $c\bar{s}$ states but with some improvement needed 
in the models \cite{Colangelo:2004vu}.  
What caught everybody's attention was how narrow these 
states were.  The problem is in the mass predictions.  Once the masses 
are fixed the narrow widths follow 
\cite{Colangelo:2004vu,Bardeen:2003kt,Godfrey:2003kg,Colangelo:2003vg}.  

The phenomenology of these states has been discussed elsewhere 
\cite{Colangelo:2004vu,Bardeen:2003kt,Godfrey:2003kg,Colangelo:2003vg}
so I 
will restrict myself to comments on some new measurements 
by Babar relating to radiative 
transitions \cite{Aubert:2006bk}.  
At the outset it was pointed out that for states this 
narrow radiative transitions are expected to have large branching 
ratios so measurement of radiative transitions is an important probe 
of their internal structure \cite{Bardeen:2003kt,Godfrey:2003kg,Colangelo:2003vg}.  
Babar obtained the following results \cite{Aubert:2006bk}:
\begin{eqnarray}
 {\cal B}(D_{sJ}(2460)^- & \to & D_s^{*-} \pi^0)  = 0.51 \pm 0.11 \pm 
0.09 \nonumber \\ 
 {\cal B}(D_{sJ}(2460)^- & \to & D_s^- \gamma) = 0.15 \pm 0.03 \pm 
0.02 \nonumber \\ 
 {\cal B}(D_{sJ}(2460)^+ & \to & D_s^+ \pi^+\pi^-) = 0.04 \pm 0.01 
\hbox{ (stat only)}  \nonumber 
\end{eqnarray}
Summing the BR's there is a missing $(30\pm 15)\%$. Where did it go?  
Recall that because $C$ is no longer a good quantum number for 
unequal mass quark and antiquark the physical $L=1$ J=1 states are a 
linear combination of $^3P_1$ and $^1P_1$ \cite{Godfrey:2003kg}:
\begin{equation}
D_{s1}^{1/2}=-^1P_1 \sin\theta + ^3P_1 \cos \theta
\end{equation}
So we expect the decay $D_{sJ}(2460)^- \to D_s^{*-} \gamma $ to occur 
and the measurement of its BR 
can be used to determine the $^3P_1-^1P_1$ mixing angle via 
\cite{Godfrey:2005ww,Godfrey:2004ct}
\begin{equation}
{ { \Gamma(^3P_1 \to ^3S_1 + \gamma)} \over  
{ \Gamma(^1P_1 \to ^1S_0 + \gamma)}}
= {{ \omega_t^3 |\langle ^3S_1 |r| ^3P_1 \rangle |^2 \cos^2\theta } 
\over
{ \omega_s^3 |\langle ^1S_0 |r| ^1P_1 \rangle |^2 \sin^2\theta } }
\end{equation}
where $\omega_t$ and $\omega_s$ are the photon energies for the two 
transitions and $\langle ^{3,1}S_{1,0} |r| ^{3,1}P_1 \rangle  $ are 
the $E1$ dipole matrix elements.  The  $1/2$ superscript refers to the 
total angular of the light quark in the heavy quark limit.

To summarize, the $D_{sJ}$ states appear to be the conventional $L=1$ 
$c\bar{s}$ states with their masses shifted due to strong $S$-wave 
coupling to $DK^{(*)}$ and their nearness to the $DK^{(*)}$ thresholds.

While almost all the theoretical effort has concentrated on the 
$D_{sJ}$ states it is important to remember that the non-strange 
partners can also provide information that can test these models
 \cite{Godfrey:1986wj,Godfrey:2005ww}.
Specifically, quark model predictions are in good agreement with the 
masses and widths of the charm $P$-wave mesons. 
Predictions for the 
radiative transitions have also been calculated.  
While the $j_q=1/2$ 
are too broad to be able to measure the radiative widths, it should be 
possible to measure the branching ratios of the
radiative transitions of the narrow states.  
In particular, measuring the BR's of the $D_1^{3/2}$ to $D\gamma$ and 
$D^*\gamma$ is a means of measuring the $^3P_1-^1P_1$ mixing angle
 \cite{Godfrey:2005ww}.

\section{$D_{sJ}(2632)$}
This state was observed by the SELEX collaboration in hadroproduction 
in $D_s^+\eta$ and $D^0K^+$ final states \cite{Evdokimov:2004iy}. 
It's measured mass is 
$M=2632.6\pm 1.6$~MeV but with the odd properties of a narrow width of 
$\Gamma <17 $~MeV at 90\% C.L. and the ratio of partial widths of 
$\Gamma(D^0K^+)/\Gamma(D^+_s\eta)=0.16\pm 0.06$.  
It has not been seen by other high statistics experiments 
\cite{Aubert:2004ku} so it's existence is in doubt.  

For the sake of argument let's investigate what it might be 
\cite{Barnes:2004ay}.  The 
possibilities mentioned in the literature are a $2^3S_1(c\bar{s})$ 
state, a $c\bar{s}$ hybrid and multiquark assignments
\cite{Maiani:2004xg,Chen:2004dy,Nicolescu:2004in}.  The lowest 
$c\bar{s}$ hybrid is expected to be about 3170~MeV so it is unlikely 
that we can identify the $D_{sJ}(2632)$ as a hybrid.  The most 
plausible conventional $c\bar{s}$ states are the $2^3S_1$ with a
predicted mass of 2730~MeV and the $1^3D_1$ with mass 2900~MeV
\cite{Barnes:2004ay}.  One 
could attribute the discrepancy with the $D_{sJ}(2632)$ mass
to mixing with the 2-meson continuum. 

If we assume the $D_{sJ}(2632)$ is the $2^3S_1(c\bar{s})$ state we can 
calculate the open-flavour decay widths and find
\begin{equation}
\Gamma(D^*K) > \Gamma (DK) >> \Gamma (D_s\eta)
\end{equation}
The total width is predicted to be $\Gamma(D_{sJ}(2632)=36$~MeV and 
$\Gamma (DK)/\Gamma (D_s\eta) \simeq 9$.  This should be compared to 
the SELEX value of $\Gamma (DK)/\Gamma (D_s\eta) =0.32\pm 0.12$.  
Clearly theory and experiment are inconsistent. It is possible to tune
the model to obtain agreement but this fine tuning seems highly 
unlikely.

We conclude that the SELEX $D_{sJ}(2632)$ needs confirmation.  
Nevertheless, experiment should be able to observe the 
$2^3S_1(c\bar{s})$ in $B$-meson decays with the largest decay mode 
predicted to be the $D^*K$ final state. 
The $1^3D_1 (c\bar{s})$ should also exist about 200~MeV higher in mass.

\section{The $X(3943)$, $Y(3943)$, and $Z(3931)$}

Three new $c\bar{c}$-like states have been observed with $C=+$.  
Their masses are consistent with the $2P$ $c\bar{c}$ multiplet and the 
$3^1S_0(c\bar{c})$ state.   Before turning to exotic interpretations 
we need to determine if they are conventional $c\bar{c}$ states.

\subsection{$X(3943)$}
The $X(3943)$ was observed by the Belle collaboration recoiling 
against $J/\psi$ in $e^+e^-$ collisions \cite{Abe:2005hd}.  
The mass and width were 
measured to be $M=3943\pm 6\pm 6$~MeV and $\Gamma =15.4\pm 10.1$~MeV.  
They find 
$BR(X\to D\bar{D}^*)=96^{+45}_{-32}\pm 22 \%$, 
$BR(X\to D\bar{D})< 41\%$ (90\% CL), 
and $BR(X\to \omega J/\psi)<26\%$ (90\% CL).  The decay to $D\bar{D}^*$ but 
not $D\bar{D}$ suggests it is an unnatural parity state.

Belle speculates that the $X(3943)$ is the $3^1S_0(c\bar{c})$ given the 
$3^3S_1(c\bar{c})$ $\psi(4040)$.  It's mass is roughly correct and the $\eta_c$ 
and $\eta_c'$ are also produced in double charm production.  This was 
also discussed by Eichten Lane and Quigg \cite{Eichten:2005ga}.  
The predicted 
width for a $3^1S_0$ with a mass of 3943~MeV is $\sim 50$~MeV 
\cite{Eichten:2005ga} which is not 
in too bad agreement with the measured 
$X(3943)$ width.  The identification of the 
$\psi(4040)$ as the $3^3S_1(c\bar{c})$ implies a hyperfine splitting 
of 88~MeV with the $X(3943)$.  This is larger than the $2S$ hyperfine 
splitting and larger than predicted by potential models. The 
discrepancy could be due to several possibilities; difficulty in fitting 
the true pole position of the $3^3S_1$ state or strong threshold
effects due to the nearby thresholds with $S$-wave and $P$-wave charm 
mesons.

The dominant $D\bar{D}^*$ final states hints at the possibility that the 
$X(3943)$ is the $2^3P_1(c\bar{c}) \; \chi_1'$ state.  It is natural 
to try the $2P(c\bar{c})$ since the $2^3P_J$ states are predicted to 
lie in the 3920-3980~MeV mass region and the widths are predicted to 
be in the range $\Gamma(2^3P_J)=30-165$~MeV \cite{Barnes:2005pb}.  
The dominant $D\bar{D}^*$ mode 
suggests that the $X(3943)$ is the $2^3P_1(c\bar{c})$ state.  The 
problems with this interpretation are that there is no evidence for the 
$1^3P_1(c\bar{c})$ state in the same data and the predicted width of 
the $2^3P_1(c\bar{c})$ is 135~MeV (assuming 
$M(2^3P_1(c\bar{c}))=3943$~MeV) \cite{Barnes}.  
Finally, there is another candidate 
for the $1^3P_1(c\bar{c})$ state, the $Y(3943)$.

To conclude, the most likely interpretation of the $X(3943)$ is that 
it is the  $3^1S_0(c\bar{c})$ $\eta_c''$ state.  
A test of this assignment is a search for this state in $\gamma\gamma 
\to D\bar{D}^*$.

\subsection{$Y(3940)$}
The $Y(3940)$ is seen by Belle in the $\omega J/\psi$ subsystem in the decay 
$B\to K\pi\pi\pi J/\psi$ \cite{Abe:2004zs}.  The reported mass and width are 
$M=3943\pm 11\pm 13$~MeV and $\Gamma=87\pm 22 \pm 26$~MeV.  It is not 
seen in $Y\to D\bar{D}$ or $D\bar{D}^*$.  The mass and width suggest a 
radially excited $P$-wave charmonium state.  But the $\omega J/\psi$ 
decay mode is peculiar.  The combined BR is 
${\cal B}(B\to KY)\cdot {\cal B}(Y\to \omega J/\psi)= (7.1\pm 1.3 \pm 
3.1) \times 10^{-5}$.  One expects that 
${\cal B}(B\to K \chi_{cJ}')<{\cal B}(B\to K \chi_{cJ})=4\times 
10^{-4}$.  This implies that ${\cal B}(Y\to \omega J/\psi)>12\%$ which 
is unusual for a $c\bar{c}$ state above open charm threshold.  

This large width to $ \omega J/\psi$ led Belle to suggest that the 
$Y(3943)$ might be a charmoniun hybrid.  The problem with this 
interpretation is that the $Y$ mass is 500~MeV below the lattice gauge 
theory estimate making the hybrid assignment unlikely.

If we identify the $Y(3940)$ with the $\chi_{c1}'$ $2^3P_1 (c\bar{c})$ 
state we expect $D\bar{D}^*$ to be the dominant decay mode with a predicted
width of 135~MeV \cite{Barnes}
which is consistent with that of the $Y(3940)$ within the 
theoretical and experimental uncertainties.  Furthermore, the 
$\chi_{c1}$ is also seen in $B$-decays. 

The decay $1^{++}\to \omega J/\psi$ is unusual.  However, the corresponding 
decay $\chi_{b1}'\to \omega \Upsilon (1S)$ has also been seen 
\cite{Severini:2003qw}.  
One possible explanation for this unusual decay mode is that rescattering 
through $D\bar{D}^*$ is responsible; 
$1^{++} \to D\bar{D}^*\to \omega J/\psi$.  
Another contributing factor might mixing with the possible 
molecular state tentatively identified with the $X(3872)$.  

We therefore tentatively identify the $Y(3940)$ as the 
 $\chi_{c1}'$ $2^3P_1 (c\bar{c})$ state.  This can be tested by 
searching for the $D\bar{D}$ and $D\bar{D}^*$ final states and by studying 
their the angular distributions ($\chi_{c1}'$ can only decay to 
$D\bar{D}^*$).

\subsection{$Z(3930)$} 
The $Z(3930)$ was observed by Belle in $\gamma \gamma \to D\bar{D}$ 
with mass and width $M=3929\pm 5\pm 2$~MeV and $\Gamma=29\pm 10 \pm 
2$~MeV \cite{Uehara:2005qd}.  The two photon width is measured to be 
$\Gamma_{\gamma\gamma}\cdot {\cal B}_{D\bar{D}}=0.18\pm 0.05\pm 
0.03$~keV.  The $D\bar{D}$ angular distribution is consistent with 
$J=2$.  It is below $D^*D^*$ threshold.

It is the obvious candidate for the $\chi_{c2}'$ $2^3P_2(c\bar{c})$ 
state.  (The $\chi_{c1}'$ cannot decay to $D\bar{D}$.)  The predicted 
mass of the $\chi_{c2}'$  is 3972~MeV.  The predicted partial widths 
and total width assuming $M(2^3P_2(c\bar{c}))=3930$~MeV are 
$\Gamma(\chi_{c2}'\to D\bar{D})=21.5$~MeV, 
$\Gamma(\chi_{c2}'\to D\bar{D}^*)=7.1$~MeV and 
$\Gamma_{total}(\chi_{c2}')=28.6$~MeV \cite{Eichten:2005ga,swanson}
in good agreement with the 
experimental measurement.  Furthermore using 
$\Gamma(\chi_{c2}'\to \gamma\gamma)= 0.67$~keV \cite{Barnes:1992sg}  
times $ {\cal B}(\chi_{c2}'\to D\bar{D})=70\%$ implies 
$\Gamma_{\gamma\gamma}\cdot {\cal B}_{D\bar{D}}=0.47$~keV which is 
within a factor of 2 of the observed number, fairly good agreement 
considering the typical reliability of 2-photon partial width 
predictions.

There is no reason to believe that the $Z(3930)$ is not the 
$\chi_{c2}'$.  However, for the sake of argument, let us consider the 
alternative possibility that it is the $\chi_{c0}'$ 
(which is not supported by the 
angular distributions).  The $\chi_{c0}'$ only decays to $D\bar{D}$ 
while the $\chi_{c2}'$ decays to both $D\bar{D}$ and $D\bar{D}^*$ in 
the ratio of $D\bar{D}^*/D\bar{D}\simeq 1/3$.  Thus, the $\chi_{c2}'$
interpretation could be confirmed by observation of the $D\bar{D}^*$
final state.  Finally we note that both the $\chi_{c2}'$  and 
$\chi_{c0}'$ undergo radiative transitions to $\psi'$ with partial 
widths $\Gamma(\chi_{c2}'\to \gamma \psi')\simeq 200$~keV and 
$\Gamma(\chi_{c0}'\to \gamma \psi')\simeq 130$~keV \cite{Barnes:2005pb}. 
Eichten Lane and Quigg find these decays are suppressed due to coupled 
channel effects \cite{Eichten:2005ga}.

\subsection{Production of $\chi'_{cJ}$ via Radiative Transitions}
It is potentially possible to observe all three $2^3P_J(c\bar{c})$ 
states in radiative decays of the $\psi(4040)$  and $\psi(4160)$ to 
$\gamma D\bar{D}$ and $\gamma D\bar{D}^*$ \cite{Barnes:2005pb}.  
The partial widths of 
$\psi(3S)\to 2^3P_J \gamma $ are 14, 39, and 54~keV for the $2^3P_2$,
$2^3P_1$, and $2^3P_0$ respectively.  Thus, all three $E1$ branching 
ratios of $\psi(4040)\to \chi_{cJ}'\gamma$ are $\sim 0.5\times 10^{-3}$.
Observing these transitions would further test whether the $X(3943)$, 
$Y(3940)$, and $Z(3930)$ are in fact the $2P(c\bar{c})$ states

\section{$X(3872)$}
The $X(3872)$ was first observed by Belle \cite{Choi:2003ue} and 
subsequently confirmed by CDF \cite{Acosta:2003zx}, D0 
\cite{Abazov:2004kp}, and Babar \cite{Aubert:2004ns}.
The mass of this state is $M=3872.0\pm 0.6\pm 0.5$~MeV and the width is 
$\Gamma < 2.3$~MeV (90 \% C.L.) which is consistent with detector 
resolution.  

This stimulated considerable speculation with a number of 
interpretations proposed in the literature; $D^0 \bar{D}^{*0}$ molecule, 
charmonium hybrid, glueball, and a conventional $2^3P_J$ or $1^3D_2$ 
state.

I'll briefly examine the possible charmonium interpretations
\cite{Barnes:2003vb,Barnes:2005pb,Eichten:2004uh,Eichten:2005ga}. Only 
the $1D$ and $2P$ multiplets are nearby in mass.  The $1^3D_1$, $1^3D_2$, 
$1^3D_3$ and $2^1P_1$ have $C=-$ (although the $\psi(3770)$ is 
identified with the $1^3D_1$) and the $1^1D_2$, $2^3P_0$,
$2^3P_1$, and $2^3P_2$ have $C=+$.  The observation of $X(3872)\to 
\gamma J/\psi$ by Belle \cite{Abe:2005ix} and Babar \cite{gowdy}
implies $C=+$.  An angular distribution analysis by the Belle 
collaboration favours $J^{PC}=1^{++}$ \cite{Abe:2005iy} although a 
higher statistics analysis by CDF cannot distinguish between 
$J^{PC}=1^{++}$ or $2^{-+}$ \cite{Kravchenko} .  Assuming it is $1^{++}$ the only surviving 
candidate is the $2^3P_1$ but as we have just seen the identification of 
the $Z(3931)$ with the $2^3P_2$ imples a $2P$ mass of $\sim 3940$~MeV 
which is inconsistent with the $2^3P_1$ interpretation.  This leads to 
the conclusion that the $X(3872)$ is a $D^0\bar{D}^{0*}$ molecule or 
``tetraquark'' state.  This is discussed in detail by Voloshin 
\cite{Voloshin:2006wf}.   
However, as just mentioned, the $2^{-+}$ is not totally ruled out 
and the predicted $2^1D_2$ mass is not too far from the observed $X(3872)$ 
mass.  A test of these hypothesis would be the observation of 
radiative transitions involving the $X(3872)$ \cite{Barnes:2003vb}.

\section{$Y(4260)$}
Perhaps the most intriguing recently discovered state is the $Y(4260)$ 
discovered by Babar as an enhancement in the $\pi\pi J/\psi$ subsystem in 
$e^+e^-\to \gamma_{ISR} J/\psi\pi\pi$ \cite{Aubert:2005rm}.  The 
measured mass and width are $M=4259\pm 8 \pm 4$~MeV and
$\Gamma =88\pm 23 \pm 5$~MeV.  The leptonic width times $BR(Y\to 
J/\psi \pi^+\pi^-)$ was measured as $\Gamma_{ee}\times BR(Y\to 
J/\psi \pi^+\pi^-)=5.5\pm 1.0 \pm 0.8$~eV.  Further evidence was seen 
by Babar in $B\to K (\pi^+\pi^- J/\psi)$ \cite{Aubert:2005zh} and 
by CLEO in $\sigma(e^+ e^-\to \pi \pi J/\psi)$ \cite{Coan:2006rv}.

The first unaccounted for $1^{--}(c\bar{c})$ state is the 
$\psi(3^3D_1)$.  Quark models estimate it's mass to be 
$M(3^3D_1)\simeq 4500$~MeV which is much too heavy to be the 
$Y(4260)$.  The $Y(4260)$ therefore represents an overpopulation of 
the expected $1^{--}$ states.  The absence of open charm production 
also argues against it being a conventional $c\bar{c}$ state.  A 
number of explanations have appeared in the literature: $\psi(4S)$ 
\cite{Llanes-Estrada:2005hz}, tetraquark \cite{Maiani:2005pe}, and 
$c\bar{c}$ hybrid \cite{Zhu:2005hp,Close:2005iz,Kou:2005gt}.

Let us consider the possibility that the $Y(4260)$ is a charmonium 
hybrid. The flux tube model predicts that the lowest $c\bar{c}$ hybrid 
mass is $\sim 4200$~MeV \cite{Barnes:1995hc} 
with lattice gauge theory having similar 
expectations \cite{Lacock:1996ny}.  Models of hybrids typically expect 
the wavefunction at the origin to vanish implying a small $e^+e^-$ 
width in agreement with the observed value.  
LGT found that the $b\bar{b}$ hybrids have large couplings to 
closed flavour models \cite{McNeile:2002az} 
which is similar to the Babar observation of 
$Y\to J/\psi \pi^+\pi^-$; the branching ratio of 
${\cal B}(Y\to J/\psi \pi^+\pi^-)>8.8\%$ combined with the observed 
width implies that $\Gamma (Y\to J/\psi \pi^+\pi^-)>7.7\pm 2.1 $~MeV.  
This is much larger than the typical charmonium transitions of, for 
example, $\Gamma(\psi(3770)\to J/\psi \pi^+\pi^-)\sim 80$~keV.  And 
the $Y$ is seen in this mode while the conventional states 
$\psi(4040)$, $\psi(4160)$, and $\psi(4415)$ are not.  

With this circumstantial evidence for the $Y(4260)$ assignment what 
measurements can be used to test this hypothesis?  
LGT suggests that we search for other closed 
charm modes with $J^{PC}=1^{--}$; $J/\psi \eta$, $J/\psi \eta'$, 
$\chi_{cJ} \omega \ldots$.  Models of hybrid decays
predict that the dominant hybrid 
charmonium open-charm decay modes will be a meson pair with an $S$-wave 
($D$, $D^*$, $D_s$, $D_s^*$) and a $P$-wave ($D_J$, $D_{sJ}$) in the 
final state \cite{Close:2005iz}.  
The dominant decay mode is expected to be $DD_1(2420)$.  
However the $D_1(2420)$ has a width of $\sim 300$~MeV and decays to 
$D^*\pi$.  This suggest the search for $Y(4260)$ in the $DD^*\pi$ 
final state.  Evidence for a large $DD_1(2420)$ signal would be strong 
evidence for the hybrid interpretation.  Having said this, it should be 
pointed out that models of hybrids have yet to be tested against 
experiment so we should be cautious.  For example, if other modes that 
were expected to be suppressed like $DD^*$ and $D_s D_s^*$ are found 
to be 
comparable to the $J/\psi \pi^+\pi^-$ mode, the $Y(4260)$ may still be 
a hybrid, but the decay models are simply not reliable.  

Another test is to search for partner states.  It is expected that the 
low lying hybrids consist of eight states in the multiplet with 
masses in the 4.0 to 4.5~GeV mass range with LGT preferring the higher 
side of the range \cite{Liao:2002rj}.  Start by confirming that no 
$c\bar{c}$ states with the same $J^{PC}$ are expected at this mass.  
Then identify $J^{PC}$ partners of the hybrid candidate which are 
nearby in mass.  It would be most convincing if some of these partners 
were found, especially the $J^{PC}$ exotics.  In the flux-tube model 
the exotic states have $J^{PC}=0^{+-}$, $1^{-+}$, and $2^{+-}$ while the 
non-exotic low lying hybrids have $0^{-+}$, $1^{+-}$, $2^{-+}$, 
$1^{++}$, and $1^{--}$.

\section{Summary} 

In the last few years there have been many new results representing 
considerable progress in our understanding of the spectroscopy 
involving charm quarks.  In some cases they have verified our models, 
in other cases they hint towards filling in missing multiplets, but 
most intriguing, in some cases they hint at non-$c\bar{c}$ states that 
could be our first evidence of qualitatively new types of hadronic 
matter. I summarize the states I discussed in the following table:

\begin{table}[h]
\begin{center}
\caption{Summary of the new charm and charmonium states discussed in 
this mini-review.}
\begin{tabular}{|l|l|}
\hline 
\textbf{State} & \textbf{Interpretation and Tests} \\
\hline 
$D_{sJ}(2317)$ & Most likely the $0^{+}(c\bar{s})$ \\
$D_{sJ}(2460)$ & Most likely the $1^{+}(c\bar{s})$ \\
$D_{sJ}(2632)$ & Needs confirmation\\
$X(3872)$ & Molecule?  see Voloshin \\
$X(3943)$ & $\eta_c''(3^1S_0)$ - look for $\gamma\gamma \to DD^*$ \\
$Y(3943)$ & $\chi_{c1}'$ - look for $D\bar{D}$ and $DD^*$ \\
$Z(3930)$ & $\chi_{c2}'$ - confirm by $DD^*$ \\
$Y(4260)$ & Hybrid? \\
\hline
\end{tabular}
\label{example_table}
\end{center}
\end{table}

To conclude I want to thank experimentalists for all the wonderful 
results they're providing!

\begin{acknowledgments}
The author thanks the organizers of FPCP'06 for their kind invitation 
and for running a wonderful meeting.  He thanks
T. Barnes, R. Faccini, 
C. Hearty, J. Rosner, and E. Swanson for helpful communications and 
discussions. This work was supported in part by the Natural Sciences 
and Engineering Research Council of Canada.
\end{acknowledgments}

\bigskip 

\end{document}